\pdfoutput=1
\documentclass[
reprint,
groupedaddress,
 amsmath,amssymb,
 aps,
 prl
]{revtex4-2}
\usepackage[table,svgnames,dvipsnames]{xcolor}
\usepackage[colorlinks=true,citecolor=blue,linkcolor=blue,urlcolor=blue, backref=false,pdfborder={0 0 0}]{hyperref}
\usepackage[normalem]{ulem}
\usepackage{aas_macros}
\usepackage{subfigure}
\usepackage{graphicx}
\usepackage{dcolumn}
\usepackage{bm}
\usepackage[mathlines]{lineno}
\usepackage{orcidlink}
\usepackage{cases}
\usepackage{booktabs}
\usepackage{comment}
\usepackage{multirow}
\usepackage{makecell}
\usepackage{siunitx}
\usepackage{booktabs}
\graphicspath{{figs/}}

\def\be{\begin{equation}}
\def\ee{\end{equation}}

\def\ba#1\ea{\begin{align*}#1\end{align*}}

\newcommand{\code}[1]{{\texttt{#1}}}
\renewcommand{\emph}[1]{\textit{#1}}

\newcommand{\refeq}[1]{Eq.~(\ref{eq:#1})}     
\newcommand{\refeqs}[2]{Eqs.~(\ref{eq:#1})--(\ref{eq:#2})}     
     
\newcommand{\reffig}[1]{Fig.~\ref{fig:#1}}

\newcommand{\reftab}[1]{Tab.~\ref{tab:#1}}

\renewcommand{\P}{\mathcal{P}}
\newcommand{\R}{\mathcal{R}}
\newcommand{\U}{\mathcal{U}}

\newcommand{\lin}{\mathrm{lin}}
\newcommand{\nonlin}{\mathrm{nonlin}}
\newcommand{\mm}{\mathrm{mm}}

\newcommand{\Om}{\Omega_m}
\newcommand{\Ob}{\Omega_b}
\newcommand{\Ok}{\Omega_K}

\newcommand{\rhob}{\bar\rho}
\renewcommand{\d}{\delta}

\newcommand{\Plin}{P^{\lin}_{\mm}}

\newcommand{\Mpch}{\,h^{-1}\text{Mpc}}

\newcommand{\hmpcinv}{\,h\,{\rm Mpc^{-1}}}
\newcommand{\kmsMpc}{\,{\rm kms^{-1}Mpc^{-1}}}

\newcommand{\vk}{\bm{k}}

\definecolor{RoyalBlue}{rgb}{0.25,.41,.88}
\definecolor{WildStrawberry}{HTML}{EE2967}

\begin{document}

\preprint{arxiv}

\title{Evidence for suppression of structure growth in the concordance cosmological model}

\author{Nhat-Minh Nguyen\orcidlink{0000-0002-2542-7233}} \email{nguyenmn@umich.edu}
\author{Dragan Huterer\orcidlink{0000-0001-6558-0112}} \email{huterer@umich.edu}
\author{Yuewei Wen\orcidlink{0000-0003-2579-7039}}
 \email{ywwen@umich.edu}
    \affiliation{Leinweber Center for Theoretical Physics, University of Michigan, 
    450 Church St, Ann Arbor, MI 48109-1040}
\affiliation{Department of Physics, College of Literature, Science and the Arts, University of Michigan, 450 Church St, Ann Arbor, MI 48109-1040}

\date{\today}

\begin{abstract}
We present evidence for a suppressed growth rate of large-scale structure during the dark-energy dominated era. Modeling the growth rate of perturbations with the ``growth index'' $\gamma$, we find that current cosmological data strongly prefer a higher growth index than the value $\gamma=0.55$ predicted by general relativity in a flat $\Lambda$CDM cosmology. Both the cosmic microwave background data from Planck and the large-scale structure data from weak lensing, galaxy clustering, and cosmic velocities separately favor growth suppression. When combined, they yield $\gamma=0.633^{+0.025}_{-0.024}$, excluding $\gamma=0.55$ at a statistical significance of 3.7$\sigma$. The combination of $f\sigma_8$ and Planck measurements prefers an even higher growth index of $\gamma=0.639^{+0.024}_{-0.025}$, corresponding to a 4.2$\sigma$-tension with the concordance model.
In Planck data, the suppressed growth rate offsets the preference for nonzero curvature and fits the data equally well as the latter model. A higher $\gamma$ leads to a higher matter fluctuation amplitude $S_8$ inferred from galaxy clustering and weak lensing measurements, and a lower $S_8$ from Planck data, effectively resolving the $S_8$ tension.
\end{abstract}

\maketitle


\textbf{Introduction.}
The flat $\Lambda$CDM concordance cosmology, which combines general relativity (GR) and a spatially flat universe with $\sim$70\% constant dark energy and $\sim$30\% cold dark matter, provides an excellent fit to observational data. However, several tensions in measurements of parameters in this model have been noted in recent years \cite{Abdalla:2022yfr}. Most significantly, the expansion rate $H_0$ inferred from the distance ladder \cite{Riess:2022} is higher than that measured by the cosmic microwave background (CMB) \cite{Planck2018:cosmology}. At a lesser significance, the parameter $S_8\equiv\sigma_8\sqrt{\Om/0.3}$ (where $\sigma_8$ is the amplitude of mass fluctuations in spheres of 8$\Mpch$ and $\Omega_m$ is matter density relative to the critical density) determined by CMB observations is larger than that found by galaxy clustering and weak gravitational lensing measurements \cite{DiValentino:2020vvd}. Finally, the Planck CMB data itself shows a preference for a nonzero spatial curvature $\Ok$ \cite{Planck2018:cosmology}.

In this Letter, we consider the possibility that the growth of structure deviates from the concordance model. While ($\Omega_m, S_8$, and $\Omega_K$) affect the growth of density perturbations, they also control geometrical quantities like distances and volumes, complicating the physical interpretation.
It is thus important to isolate and constrain the growth of structure \cite{Huterer:2022dds} separately from geometrical quantities. Here, we adopt a precise parameterization of the growth rate and find evidence for growth suppression---relative to the expectation from flat $\Lambda$CDM and GR---which also reconciles tensions in $S_8$ and $\Ok$ constraints.
Our results clarify and consolidate the current situation in the field, where different analyses adopting different prescriptions of growth (and geometry), either found some evidence for a suppressed growth \cite{Ruiz-Huterer:2014,Bernal:2015,Johnson:2016,Moresco:2017hwt,Basilakos:2020,Said:2020,Garcia:2021,Ruiz-Zapatero:2022,White:2022,Chen:2022,DESY3:extension2022} or did not \cite{Wang:2007,Guzzo:2008ac,Dossett:2010gq,Hudson:2012,Rapetti:2012bu,Pouri:2014nta,Alam:2016,Ruiz-Zapatero:2021,Muir:2021,Andrade:2021}.
Our baseline constraint is consistent with \cite{Dossett:2010gq,Hudson:2012,Rapetti:2012bu,Pouri:2014nta,Alam:2016}, whose constraints are also consistent with standard growth rate in $\Lambda$CDM and GR given their data and modeling uncertainties.

\medskip
\textbf{Growth of structure.}
Over cosmic time, matter density fluctuations $\d\equiv (\rho-\rhob)/\rhob$ (where $\rho$ and $\rhob$ are the local and the cosmic mean densities respectively) are amplified by gravity.
Assuming GR and restricting to linear regime where $\d\ll1$ ($k\lesssim 0.1\hmpcinv$ today with $h = H_0/100\kmsMpc$) and subhorizon scales ($k\gtrsim H_0\simeq 0.0003\hmpcinv$ today), we can describe the growth of large-scale structure as \cite{Peebles:lss,Bernardeau:lss_review}
\be
\ddot\d(\vk,t) + 2H\dot\d(\vk,t) - 4\pi G\bar\rho \d(\vk,t) = 0, 
\label{eq:growth_ODE}
\ee
where dot denotes derivative with respect to time; the matter overdensity $\d$, the expansion rate $H$, and the mean matter density $\bar\rho$ all depend on time, while every Fourier $\vk$-mode evolves independently.
Linear growth is thereby described by the linear growth function $D(t)\equiv \d(t)/\d(t_0)$, where $t_0$ denotes the present, and the growth rate $f(a)\equiv d\ln D(a)/d\ln a$, where $a(t)$ is the scale factor.
The growth rate is a central link between data and theory: it is proportional to large-scale structure observables like peculiar velocities and redshift-space distortions \cite{Peebles:1976,Lightman_Schechter:1990}, while being exquisitely sensitive to the properties of dark-energy models \cite{Cooray:2003hd}.

To isolate the temporal evolution of structure, \cite{Fry:1985,Wang_Steinhardt:1998,Linder:2005} introduced a robust and accurate approximation of the growth rate as
\be
f(a)=\Om^{\gamma}(a),
\label{eq:gamma_parametrization}
\ee
where $\gamma$ is the \emph{growth index}.
In particular, \cite{Wang_Steinhardt:1998,Linder:2005} showed that standard GR in the flat $\Lambda$CDM background predicts $\gamma\simeq0.55$ even in the presence of dark energy; this fit is accurate to $\simeq 0.1\%$ \cite{Linder:2005,Linder_Cahn:2007,Gong:2008}. A measured deviation from $\gamma=0.55$ would suggest an inconsistency between the concordance cosmological model and observations.

Assuming \refeq{gamma_parametrization}, the linear growth function takes the form
\be
  \label{eq:D_gamma}
  D(\gamma,a)=\exp\left[-\int_a^1\,da\,\frac{\Om^{\gamma}(a)}{a}\right],
\ee
where we have normalized $D(\gamma,a=1)\equiv 1$ for all $\gamma$. A $\gamma>0.55$ corresponds to a growth rate $f(\gamma,a)<f(0.55,a)$ and, for our present-day normalization, to a growth function $D(\gamma,a)>D(0.55,a)$ in the past.

\medskip
\textbf{Methodology and data.}
To implement \refeqs{gamma_parametrization}{D_gamma}, we express the linear matter power spectrum as
\be
  \label{eq:scaling_plin}
  P(\gamma,k,a)=P_{\mathrm{today}}(k,a=1)\,D^2(\gamma,a),
\ee
where $P_{\mathrm{today}}$ is the fiducial linear matter power spectrum evaluated today which depends on the usual set of cosmological parameters.
We note that the choice of $a(t)$ at which growth is normalized does not impact our $\gamma$ constraints and its (in)consistency with $\gamma=0.55$ as we jointly infer the power spectrum amplitude as well (see below).
To compute transfer functions and power spectra, we modify the cosmological Boltzmann solver \code{CAMB} \cite{Lewis:camb,Howlett:camb}.
With $\gamma=0.55$ we obtain (at redshift $z=1.5$ and up to $k\lesssim 0.1\hmpcinv$) linear matter power spectra within 0.1\% of the outputs from the unmodified version of \code{CAMB}. Likewise, we repeat the baseline Planck 2018 \cite{Planck2018:cosmology} and DES year-1 \cite{DESY1:3x2pt} analyses, using our modified \code{CAMB} \footnote{Code available at this fork of \code{CAMB}: \href{https://github.com/MinhMPA/CAMB_GammaPrime_Growth}{{github.com/MinhMPA/CAMB\_GammaPrime\_Growth}}.} at fixed $\gamma=0.55$, and reproduce their constraints on relevant cosmological parameters well within their precision.

Because the growth-index parameterization has only been validated for sub-horizon perturbations, care needs to be taken when modeling the CMB whose information partially comes from large scales and high redshifts.
Therefore, we isolate the effect of $\gamma$ from the prediction for the (unlensed) primary CMB anisotropies. \refeq{scaling_plin} only modifies the CMB lensing gravitational potential \footnote{The integrated Sachs-Wolfe effect \cite{Sachs:1967er,Carron:2022eum}, a secondary CMB anisotropy sourced by gravitational redshift, is also affected by $\gamma$ and \refeq{scaling_plin}. We do not consider that signal here.}, which is generated by density fluctuations within the regime where \refeqs{gamma_parametrization}{scaling_plin} are valid. 

Our baseline data includes measurements of the parameter combination $f\sigma_8$ from peculiar velocity and redshift-space distortion (RSD) data, at local ($z<0.1$) \cite{6dF:growth_sigma8,Huterer:2016,Said:2020,Boruah:2020,Turner:2023} and cosmological distances ($z\geq0.1$) \cite{Blake:2011,Blake:2013,Howlett:2015,Okumura:2016,Pezzotta:2017,SDSS:DR16}. \reffig{fsigma8z_constraints_vs_data} shows these $f\sigma_8$ measurements at the corresponding redshifts. We assume that the $f\sigma_8$ measurement uncertainties are Gaussian-distributed and uncorrelated among each other \footnote{Likelihood and data available at: \href{https://github.com/MinhMPA/cobaya}{{github.com/MinhMPA/cobaya}}}. We further complement the $f\sigma_8$ measurements with either the Planck 2018 CMB data---including CMB temperature-temperature, temperature-polarization plus polarization-polarization spectra and CMB lensing reconstruction \cite{Planck2018:cosmology,Planck2018:lensing,Planck2018:likelihood} (hereafter PL18 collectively)---or large-scale structure data from galaxy surveys, or both. Data from galaxy surveys include a) the DESY1 3x2pt correlation functions \cite{DESY1:3x2pt} (hereafter DESY1), and b) baryon acoustic oscillations in the 6dF Galaxy Survey (6dFGS) galaxy \cite{6dF:BAO_H0} and the Sloan Digital Sky Survey (SDSS) \cite{SDSS:DR7,SDSS:DR12,SDSS:DR16} galaxy plus Lyman-alpha (hereafter BAO collectively). When including both SDSS $f\sigma_8$ and BAO data, we employ joint covariance and likelihood that properly account for their correlations \footnote{\href{https://cobaya.readthedocs.io/en/latest/likelihood_bao.html}{cobaya.readthedocs.io/en/latest/likelihood\_bao.html}}.
Throughout, we adopt the same likelihoods and priors used in the baseline of those analyses. We fix the total mass of neutrinos to $\sum m_\nu=0.06$ eV and include neutrino contribution $\Omega_\nu$ in the matter density parameter $\Om$. We verify that excluding $\Omega_\nu$ in computing theoretical $f\sigma_8$ leads to negligible changes in the latter and all downstream results. We allow $\gamma$ to vary assuming a uniform prior $\U(0,2.0)$.

We constrain the growth index $\gamma$, along with other standard cosmological parameters: the matter and baryon densities relative to critical $\Om$ and $\Ob$, the Hubble constant $H_0$, spectral index $n_s$, mass fluctuation amplitude $\sigma_8$, and reionization optical depth $\tau$.
We therefore perform Bayesian inference via the Monte Carlo Markov Chain (MCMC) method using the \code{cobaya} framework \cite{Torrado_Lewis:cobaya} and analyze the MCMC samples using the \code{GetDist} package \cite{Lewis:getdist}.

To quantify the statistical significance of our results, we compute the Bayesian factor of $\gamma=0.55$ and $\gamma\neq0.55$ by assuming the Savage-Dickey density ratio
\be
\log_{10}\mathrm{BF}_{01}=\log_{10}\left.\frac{\P(\gamma|\mathrm{d,M_1})}{\P(\gamma|\mathrm{M_1})}\right|_{\gamma=0.55},
\label{eq:Bayes_factor}
\ee
where $\mathrm{d}$ and $\mathrm{M_1}$ respectively denote the data and the model with $\gamma$, while $\P(\gamma|\mathrm{M_1})=\U(0.,2.)$. This is reported in the fifth column of \reftab{results}.
We further quote the significance of $\gamma\neq0.55$ following the two-tailed test and measuring the posterior tail in units of Gaussian sigmas.
In the Supplementary Material, we compare the goodness-of-fit of models with respect to each data combination, and for each individual likelihood.

\begin{figure}[!t]
   \centering
   \includegraphics[width=\linewidth]{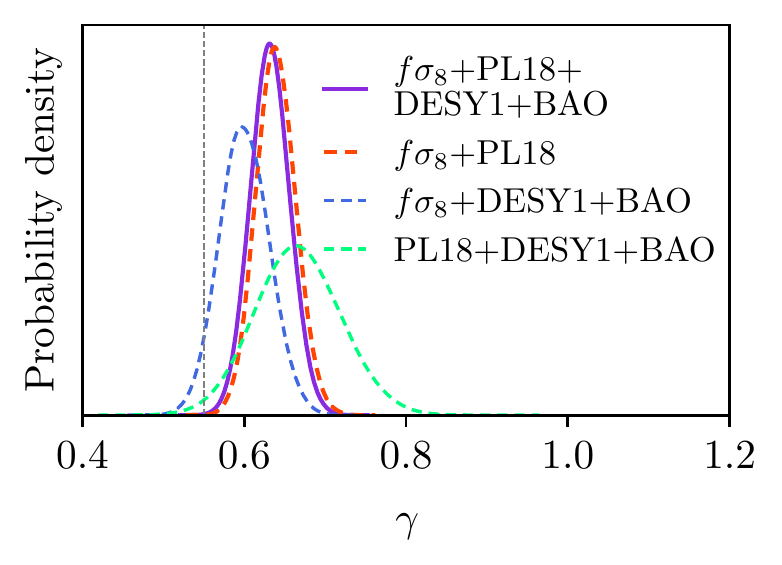}
   \caption{Marginalized constraints on the growth index $\gamma$, from CMB (PL18) and LSS data. The latter includes $f\sigma_8$, DES-Y1 and BAO measurements. Legend indicates different combinations of the data sets.} The vertical dashed line marks the concordance model prediction of $\gamma=0.55$.
   \label{fig:gamma_1Dposteriors}
 \end{figure}

\begin{figure}[!t]
   \centering
   \includegraphics[width=\linewidth]{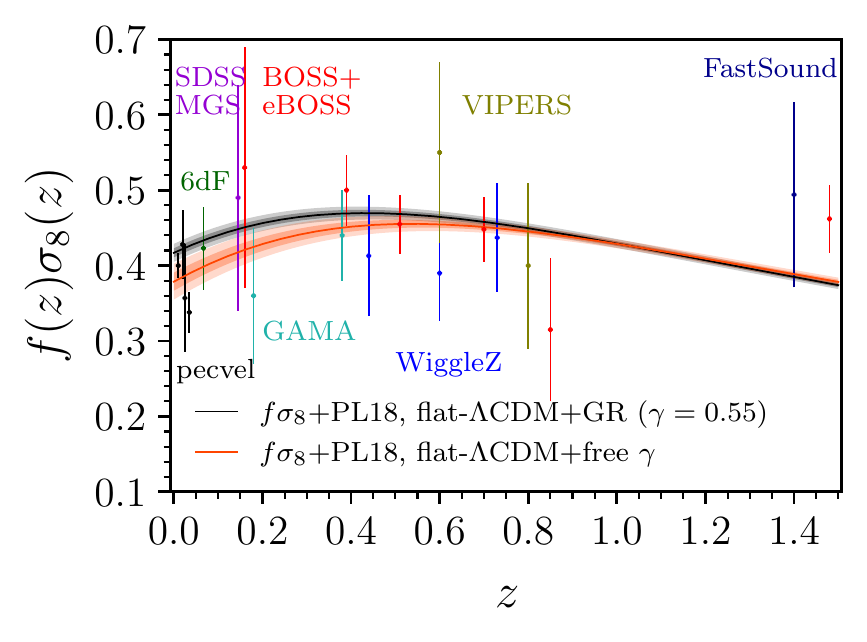}
   \caption{Marginalized posterior on the theoretical $f(z)\sigma_8(z)$ assuming the growth-index parameterization in \refeq{gamma_parametrization}. Shaded bands show the 68\% and 95\% posteriors from our baseline analysis that includes $f\sigma_8$ and PL18 data (orange), and the corresponding constraints in the concordance model with $\gamma=0.55$ (black). The data points indicate actual $f\sigma_8$ measurements.}
   \label{fig:fsigma8z_constraints_vs_data}
 \end{figure}

\medskip
\textbf{Constraints on $\gamma$ in a flat universe.}
We first consider the data combination $f\sigma_8$+PL18. 
Marginalizing over all other cosmological parameters, we obtain the orange posterior density in \reffig{gamma_1Dposteriors}.
 This corresponds to the constraint $\gamma=0.639^{+0.024}_{-0.025}$ and a Bayes factor of $|\log_{10}\mathrm{BF}_{01}|=1.7$. The former excludes $\gamma=0.55$ at a statistical significance of 4.2$\sigma$, while the latter provides a ``very strong'' evidence for deviation from the GR+flat $\Lambda$CDM prediction of $\gamma=0.55$ according to the Jeffreys' scale \cite{Jeffreys:probability_theory}.
Neither PL18 nor $f\sigma_8$ alone substantially constrains the growth index due to degeneracies with other cosmological parameters, yet together they show a clear preference for $\gamma>0.55$, that is, a lower rate of growth than predicted by GR in flat $\Lambda$CDM.
\reffig{fsigma8z_constraints_vs_data} illustrates the effect of growth suppression as a function of redshift by showing the $f(z)\sigma_8(z)$ posterior assuming flat $\Lambda$CDM, and that assuming flat $\Lambda$CDM+$\gamma$, both inferred from the $f\sigma_8$+PL18 data combination.
 
Next, we investigate how the galaxy clustering and lensing data constrain $\gamma$.
To do so, we replace the PL18 data by the DESY1 3x2pt measurements of galaxy clustering and weak lensing, together with the expansion-history data from BAO. The $f\sigma_8$+DESY1+BAO data combination yields the marginalized constraint $\gamma=0.598^{+0.031}_{-0.031}$. Much like the $f\sigma_8$ + PL18 constraint, this combination prefers a higher growth index than the GR value, except now at a lower statistical significance, excluding $\gamma=0.55$ at 2.0$\sigma$.
 
We finally report the constraint from all data combined, $f\sigma_8$+PL18+DESY1+BAO:
\be
\gamma = 0.633^{+0.025}_{-0.024}.
\label{eq:fsigma8+PL18+DESY1+BAO_gamma_constraint}
\ee
Analysis of the posterior tails indicates that $\gamma=0.55$ is excluded at 3.7$\sigma$, while the Bayes factor $|\log_{10}\mathrm{BF}|=1.2$ shows a ``strong'' evidence for a departure from the expected value of $\gamma$.
The constraint is represented by the violet posterior density in \reffig{gamma_1Dposteriors};
it is very close to the posterior for $f\sigma_8$+PL18.
For clarity, we additionally plot $\gamma$ constraint from PL18+DESY1+BAO in green.

We summarize all $\gamma$ constraints, together with their statistical significance, in \reftab{results}. We further assert the robustness of and internal consistency between our $\gamma$ constraints in the Supplementary Material.
 
\begin{figure*}[th]
   \centering
   \includegraphics[width=\linewidth]{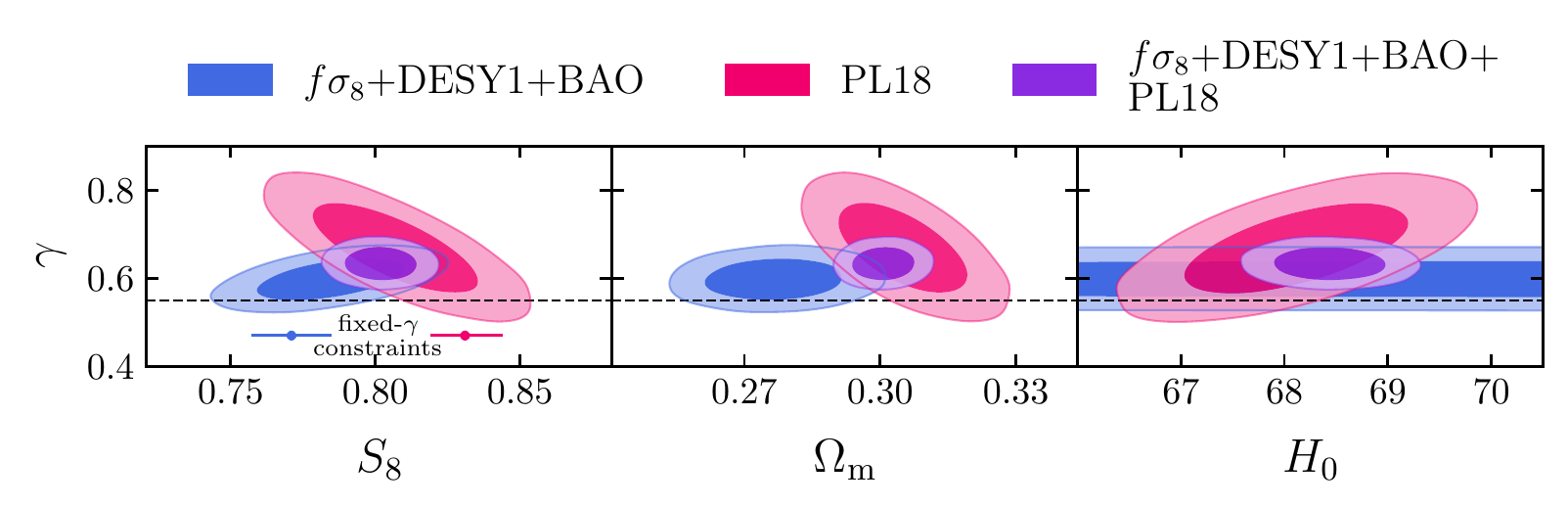}
   \caption{68\% and 95\% marginalized constraints on parameters in the concordance model allowing for a free growth index $\gamma$, from $f\sigma_8$+DESY1+BAO (blue), PL18 alone (red) and $f\sigma_8$+DESY1+BAO+PL18 (violet). Contours contain 68\% and 95\% of the corresponding projected 2D constraints. The horizontal black dashed lines mark the concordance model prediction of $\gamma=0.55$. The horizontal bars in the $\gamma-S_8$ panel indicate the 68\% limits on $S_8$ for a fixed $\gamma=0.55$ (see text); they are vertically offset from $\gamma=0.55$ for visibility.}
   \label{fig:gamma_S8_omegam_H0_2Dcontours}
 \end{figure*}
 
\medskip
\textbf{Implications for $S_8$ tension.}
A moderate yet persistent tension in constraints of $S_8$ has emerged between CMB measurements, e.g.\ Planck \cite{Planck2018:cosmology} or Atacama Cosmology Telescope plus Wilkinson Microwave Anisotropy Probe \cite{ACTDR4:cosmology}, and low-redshift 3x2pt measurements of weak lensing and galaxy clustering, e.g. the Dark Energy Survey (DES) \cite{DESY1:3x2pt}, the Kilo-Degree Survey (KiDS) \cite{KIDS-1000:3x2pt}, and combinations thereof \cite{Amon:2023}.
This discrepancy is statistically significant and unlikely to be explained by lensing systematics alone \cite{Lensing_wo_borders}, thus motivates investigations of physics beyond the standard model.

\reffig{gamma_S8_omegam_H0_2Dcontours} shows the marginalized constraints in the 2D planes of the growth index $\gamma$ and, from left to right, $S_8$ or $\Omega_m$ or $H_0$, by different data combinations. Notably, the $S_8-\gamma$ panel indicates a potential solution to the $S_8$ tension: a higher growth index ($\gamma\simeq 0.65$) implies a \emph{higher} $S_8$ value in the probes of large-scale structure. Specifically, the $f\sigma_8$+DESY1+BAO combination yields $S_8=0.784^{+0.017}_{-0.016}$, while in the standard $\Lambda$CDM (with $\gamma\equiv 0.55$) $S_8=0.771^{+0.014}_{-0.014}$. Conversely, Planck now prefers a \emph{lower} amplitude of fluctuations ($S_8=0.807^{+0.019}_{-0.019}$) than it does in $\Lambda$CDM ($S_8=0.831^{+0.013}_{-0.012}$). Consequently, the ``$S_8$ tension'' between the measurements of $S_8$ in the galaxy clustering and gravitational lensing versus that in Planck decreases from $3.2\sigma$ to 0.9$\sigma$, as measured by the $S_8$ difference divided by errors added in quadrature.

\begin{table*}[htp]
  \caption{Constraints on the growth index $\gamma$ and cosmological parameters $S_8$ and $H_0$ from different data combinations, the corresponding Bayes factors, and chi-square differences relative to the concordance model ($\gamma=0.55$).}
  \begin{center}
  	\begin{ruledtabular}
    	\begin{tabular}{lc c c c c c}
      		Data & $\gamma$ & $S_8$ & $H_0$ [$\kmsMpc$] & $|\log_{10}\mathrm{BF}_{10}|$ & $\Delta\chi^2\equiv \chi^2_{\gamma}-\chi^2_{\gamma=0.55}$\\
      		\colrule
      		PL18 & $\mathbf{0.668^{+0.068}_{-0.067}}$ & $0.807^{+0.019}_{-0.019}$ & $68.1^{+0.7}_{-0.7}$ & $0.4$ & $-2.8$\\
     		PL18+$f\sigma_8$ & $\mathbf{0.639^{+0.024}_{-0.025}}$ & $0.814^{+0.011}_{-0.011}$ & $67.9^{+0.5}_{-0.5}$ & $1.7$ & $-13.6$\\
     		PL18+$f\sigma_8$+DESY1+BAO & $\mathbf{0.633^{+0.025}_{-0.024}}$ & $0.802^{+0.008}_{-0.008}$ & $68.4^{+0.4}_{-0.4}$ & $1.2$ & $-13.2$\\
        \colrule
     PL18+$f\sigma_8$+DESY1+BAO (flat $\Lambda$CDM+GR) & $\mathbf{0.55}$ & $0.803^{+0.008}_{-0.008}$ & $68.5^{+0.4}_{-0.4}$ & - & 0
     	\end{tabular}
     \end{ruledtabular}
  \end{center}
  \label{tab:results}
\end{table*}

\medskip
\textbf{Allowing curvature to vary.}
Relaxing the assumption of spatial flatness changes the expansion history and the concordance prediction for the growth history \cite{Mortonson:2009,Gong:2009}. An immediate question is whether the apparent preference for a higher growth index and a slower growth rate is the same effect as the apparent preference for a nonzero curvature found by the Planck 2018 analysis that, by using temperature and polarization data, found $\Ok=-0.044^{+0.018}_{-0.015}$ (\cite{Planck2018:cosmology}; see also \cite{DiValentino:2020,Handley:2021,DiValentino:2022}). 

\begin{figure}[t]
   \centering
   \includegraphics[width=\linewidth]{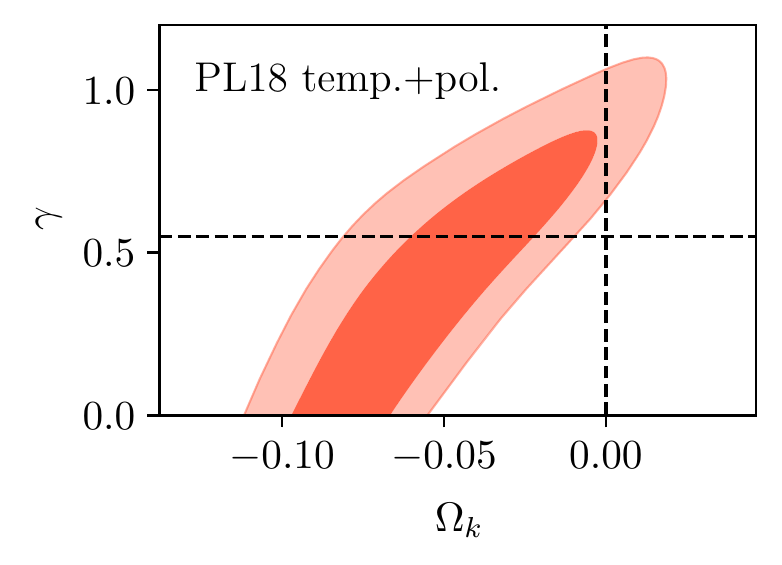}
   \caption{Degeneracy between $\gamma$ and $\Ok$ in the PL18 temp.+pol. analysis when both parameters are allowed to vary. Contours show the 68\% and 95\% credible intervals. The dashed lines mark the point $[\Ok=0,\gamma=0.55]$ corresponding to the concordance flat $\Lambda$CDM model.}
   \label{fig:gamma_omegak_2Dcontours}
 \end{figure}

Allowing both curvature and growth index to vary, we observe a trade-off between $\Ok$ and $\gamma$, as shown in \reffig{gamma_omegak_2Dcontours} using \emph{only} Planck CMB temperature and polarization data (henceforth PL18 temp.+pol.). The data clearly prefer either a positively curved space, i.e.\ $\Ok<0$, or growth suppressed relative to the GR prediction, i.e.\ $\gamma>0.55$; the flat model with $\gamma=0.55$ has a worse fit than the best-fit model by $\Delta\chi^2=-6.9$.

We next focus on two limits of the results shown in \reffig{gamma_omegak_2Dcontours}: a) varying $\Ok$ while fixing $\gamma=0.55$ 
(which reproduces the standard analysis from the Planck paper, also finding $\Ok=-0.044$), and b) fixing $\Ok=0$ while varying $\gamma$. We are particularly interested in comparing the fit of these two models. We find that the model with free curvature fits the PL18 temp.+pol. data marginally better than the model with free $\gamma$ ($\Delta\chi^2=-1.3$). Including PL18 CMB lensing reconstruction likelihood leads to $\Delta\chi^2=0.7$ in favor of the free-$\gamma$ model. Overall, we conclude that both models fit the PL18 data equally well.

\begin{figure}[t]
   \centering
 \includegraphics[width=\linewidth]{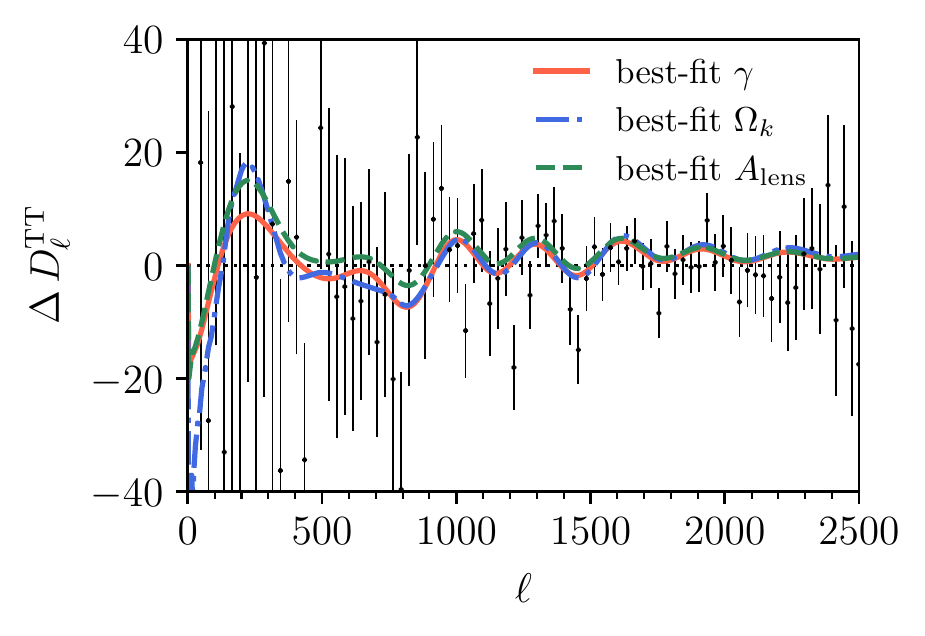}
   \caption{Residuals in the CMB TT power spectrum $D_{\ell}\equiv\ell(\ell+1)C_\ell/(2\pi)$ between the best-fit model with free $\gamma$ (orange), best-fit model with curvature (blue), and best-fit model with free CMB lensing amplitude $A_{\mathrm{lens}}$ (green). The data points and error bars represent the Planck 2018 (binned) TT power spectrum residuals and the 68\% uncertainties. All residuals are computed with respect to the best-fit concordance model.}
   \label{fig:gamma_vs_curvature_vs_alens_bestfit_DellTT_residuals}
 \end{figure}

Recall that the feature in the PL18 temp.+pol. data driving the preference for $\Ok<0$ is essentially the same one that favors a high CMB lensing amplitude, i.e.\ $A_\mathrm{lens}>1$ \cite{Planck2015:cosmology,Planck2018:cosmology,Planck2018:likelihood}. Does the cosmological model with a high $\gamma$ produce similar features in the CMB power spectra as those with $\Ok<0$ or $A_\mathrm{lens}>1$? The answer is affirmative, as shown in \reffig{gamma_vs_curvature_vs_alens_bestfit_DellTT_residuals} where we compare the residuals in the CMB temperature power spectrum (TT) of a) the PL18 data, b) the best-fit flat model with $\gamma$, c) the best-fit model with curvature but fixed $\gamma=0.55$, and d) the best-fit flat model with $A_\mathrm{lens}$ but fixed $\gamma=0.55$, all relative to that of the best-fit concordance model. All three best-fit model residuals display the same oscillatory pattern that closely follows the oscillations in the data residuals.
The similarity between the best-fit models with $\gamma>0.55$ (case b) and with $A_\mathrm{lens}>1$ (case d) in the CMB power spectra is not entirely surprising: a higher $\gamma$ encodes a lower growth rate $f(a)$ and, for a fixed amount of structure observed today, a higher growth (relative to standard growth $\gamma=0.55$) in the recent past (see \refeq{D_gamma}). This in turn implies a higher lensing amplitude, thus has a qualitatively similar effect as $A_\mathrm{lens}>1$. We illustrate the effect of $\gamma$ on the lensing potential power spectrum in the Supplementary Material \footnote{The preference for anomalous growth index in PL18 temp.+pol. data ($\Delta\chi^2=-8.5$ in favor of the free-$\gamma$ model over the concordance one) decreases once the CMB lensing reconstruction likelihood is included ($\Delta\chi^2=-2.8$). A similar effect is observed for the case of varying $A_\mathrm{lens}$.}.

\medskip
\textbf{Summary and Discussion.}
In this Letter, we have presented new constraints on the growth rate using a combination of Planck, DES, BAO, redshift-space distortion and peculiar velocity measurements. The constraints from different data combinations are consistent with one another within 1$\sigma$. Our constraints exclude the predictions of flat $\Lambda$CDM model in GR at the statistical significance of 3.7$\sigma$, indicating a suppression of growth rate during the dark-energy dominated epoch.

Further, we have demonstrated that cosmological models with a high $\gamma$ resolve two known tensions in cosmology. First, allowing for a suppressed growth removes the need for negative curvature indicated by the PL18 temp.+pol. data; in fact, the best-fit flat model with free $\gamma$ fits the data equally well as the best-fit model with standard growth and negative curvature, producing highly similar features in the temperature power spectrum. Second, the discrepancy in the measured amplitude of mass fluctuations parameter $S_8$ from the PL18 data and that from the large-scale structure data can be reconciled with a high-$\gamma$ model. Our findings indicate that these cosmological tensions can be interpreted as evidence of growth suppression.

A late-time linear growth suppression is not straightforward to achieve in modified theories of gravity, particularly if the expansion history is similar to that in the concordance model \cite{Barreira:2012,Joyce:2016,Kable:2022} as our constraints indicate. Nevertheless, there is sufficient freedom in the space of modified-gravity theory (within a sub-class of Horndeski models, e.g. \cite{Piazza:2014,Perenon:2015,Perenon:2019dpc,Wen:2023bcj}) to do so. \footnote{A scale-dependent suppression would open up more possibilities, e.g. alternative models of dark matter \cite{FrancoAbellan:2020xnr,Rogers:2023ezo}.} Probing such modified-gravity theories should be within the reach of future surveys and experiments \cite{Frusciante:2019,Perenon:2020,Wen:2023bcj}. Specifically, upcoming large-scale structure data \cite{Taipan:whitepaper2017,HETDEX:whitepaper2021,DESI:roadmap2022,PFS:whitepaper2014,Euclid:whitepaper2011,MegaMapper:whitepaper2022} will improve $f\sigma_8$ data both in terms of measurement precision and redshift coverage. In parallel, forthcoming CMB measurements \cite{ACTDR4:cosmology,SimonsObservatory:science2019,CMB-S4:science2016,LiteBIRD:2022} with higher resolution and sensitivity will play a significant role in pinning down the expansion history and growth rate.
In this era of high-precision large-scale structure and CMB measurements, joint analyses of these data sets will hold the key to confirming any evidence for physics beyond the standard model.

\medskip
\begin{acknowledgments}
We are grateful to Eiichiro Komatsu, Eric Linder, Jessie Muir, Fabian Schmidt and the three anonymous referees for their valuable comments on the manuscript. MN thanks Alex Mead for helpful conversations on details and modifications of \code{HMcode-2020}. We thank Alex Barreira, Elisa Ferreira, Shaun Hotchkiss, Jiamin Hou, Cullan Howlett, Mike Hudson, St\'ephane Ili\'c, Johannes Lange, and Antony Lewis for useful discussions.
We acknowledge support from the Leinweber Center for Theoretical Physics, NASA grant under contract 19-ATP19-0058, DOE under contract DE-FG02-95ER40899, and the University of Michigan Research Computing Package. Our analysis was performed on the Greatlakes HPC cluster, maintained by the Advanced Research Computing division, UofM Information and Technology Service. MN thanks John Thiels and Mark Champe for going above and beyond during their service.
This work was initiated at the Aspen Center for Physics, which was supported by the National Science Foundation grant PHY-1607611. We thank the Aspen Center for their hospitality.
\end{acknowledgments}

\bibliographystyle{apsrev4-2}
\bibliography{references}

\newpage
\pagebreak

\appendix
\onecolumngrid

\section{\Large Supplementary Material}

\section{\texorpdfstring{I.\hspace{0.05in}Individual best-fit $\Delta\chi^2$}{Individual best-fit delta chi-squares}}\label{subsec:delta_chisquares}
In \reftab{individual_delta_chisquares}, we compare the goodness-of-fit between best-fit models with free $\gamma$ and those with $\gamma=0.55$ for different data sets and combinations. We identify best-fit models as those that maximize the corresponding joint posteriors. In general, best-fit models with $\gamma>0.55$ tend to fit individual data set and likelihood better, as indicated by the corresponding negative $\Delta\chi^2$, with the notable exception of the PL18 CMB lensing case. We will return to this case in a later section.

\begin{table*}[htp]
  \caption{Chi-square differences between best-fit models with free $\gamma$ and best-fit concordance models, for different data combinations and individual likelihoods. ``TT'', ``EE'' refer to PL18 temperature and polarization auto power spectra respectively, while ``TTTEEE'' refers to both the auto and cross power spectra.}
  \begin{center}
          \begin{ruledtabular}
    	\begin{tabular}{l c c c c c c c c}
      		Data & \multicolumn{8}{c}{$\Delta\chi^2\equiv \chi^2_{\gamma}-\chi^2_{\gamma=0.55}$}\\
                \cline{2-9}
                 & \textbf{low-$\ell$ TT} & \textbf{low-$\ell$ EE} & \textbf{high-$\ell$ TTTEEE} & \thead{\textbf{lensing}\\\textbf{reconstruction}} & \textbf{$f\sigma_8$} & \textbf{DESY1} & \textbf{BAO} & \textbf{total}\\
                 PL18 temp.+pol. &$-1.1$ &$-0.4$ &$-7.0$ &- &- &- &- &$-8.5$\\ 
                 PL18 &$-1.0$ &$-0.1$ &$-3.1$ &$+1.4$ &- &- &- &$-2.8$\\
                 PL18$+f\sigma_8$ &$+0.1$ &$-0.3$ &$-5.6$ &$+0.5$ &$-8.3$ &- &- &$-13.6$\\
                 PL18$+$DESY1$+$BAO &$-0.6$ &$-0.8$ &$-3.7$ &$+0.3$ &- &$-0.7$ &$+0.8$ &$-4.7$\\
                 $f\sigma_8+$DESY1$+$BAO &- &- &- &- &$-1.2$ &$-2.9$ &$-2.2$ &$-6.3$\\
                 PL18$+f\sigma_8+$DESY1$+$BAO &$-0.2$ &$-1.1$ &$-5.3$ &$-0.7$ &$-6.8$ &$+0.8$ &$+0.1$ &$-13.2$\\
     	\end{tabular}
        \end{ruledtabular}
  \end{center}
  \label{tab:individual_delta_chisquares}
\end{table*}

\section{II.\hspace{0.05in}Modeling of nonlinear power spectrum}\label{subsec:nonlinear-pk}

Both PL18 \cite{Planck2018:cosmology} and DESY1 \cite{DESY1:3x2pt} official analyses modelled the matter power spectrum into the nonlinear regime, using different approaches but based on the same halo model framework \cite{Seljak:2000gq,Ma:2000ik,Peacock:2000qk}. The framework approximates the nonlinear matter distribution under the assumption that all matter resides in halos, hence matter clustering can be partitioned into inter- (two-halo) and intra-halo (one-halo) regimes. The nonlinear matter power spectrum $P^{\nonlin}_{\mm}(\gamma,k,a)$ can then be written as a sum of these two terms:
\be
  \label{eq:nonlin_pk}
  P^{\nonlin}_{\mm}(\gamma,k,a)=P^{2\mathrm{H}}_{\mm}(\gamma,k,a)+P^{1\mathrm{H}}_{\mm}(\gamma,k,a),
\ee
where both terms are integrated over the entire halo mass range $M\in[0,\infty)$.
The two-halo term $P^{2\mathrm{H}}_{\mm}(\gamma,k,a)=\Plin(\gamma,k,a)\left[\int_0^\infty \,b(M,a)W(M,a,k)n(M,a)dM\right]^2$ describes matter cluttering between two distinct halos, hence the explicit relation to the linear matter power spectrum in Eq.~(4) of the Letter.
The one-halo term $P^{1\mathrm{H}}_{\mm}(\gamma,k,a)=\int_0^\infty\,W^2(M,k,a)n(M,a)dM$ describes matter clustering within individual halos.
Here, the linear matter power spectrum in Eq.~(4) of the Letter enters only through the variance of the matter field inside the halo number density $n(M,a)$ \cite{Sheth:1999mn,Jenkins:2000bv}.
In the above expressions, $b$ and $W$ are the halo bias factor and the halo density profile, respectively.

In this work, similar to \cite{Planck2018:cosmology,Kilo-DegreeSurvey:2023gfr}, we follow the \code{HMCode-2020} approach \cite{Mead:hmcode-2020} for the connection in \refeq{nonlin_pk}. Specifically, we adopt the \code{HMCode-2020} version implemented in \code{CAMB} nonlinear module \href{https://github.com/MinhMPA/CAMB_GammaPrime_Growth/blob/master/fortran/halofit.f90}{halofit.f90}. The code augments the halo model with parameters that modify $P^{2\mathrm{H}}_{\mm}$ and $P^{1\mathrm{H}}_{\mm}$ to account for the effects of nonlinear gravitational evolution and baryonic feedback. The parameters were calibrated by fitting model prediction of $P^{\nonlin}_{\mm}$ to emulated and simulated data \cite{Mead:hmcode-2020}.
For consistency, we further modify the linear growth function in \code{HMCode-2020} such that $D(a)\to D(\gamma,a)$. We set all other model parameters in \code{HMCode-2020} to their default values.

We note that our conclusions do not depend on the details of the modeling choices by \code{HMCode-2020}, as we have verified that the PL18 and DES-Y1 fiducial cosmological constraints can be reproduced with $\gamma=0.55$. In other words, if the data prefer expansion and growth histories consistent with GR+flat $\Lambda$CDM, $\gamma$ must be consistent with 0.55.

\section{\texorpdfstring{III.\hspace{0.05in}$\gamma$ and CMB lensing potential}{Gamma and CMB lensing potential}}\label{subsec:CMBlensing}
For simplicity, below we assume a) the Born approximation, b) the gravitational potential is the same in the Weyl and Newtonian gauges and c) the recombination epoch was instantaneous such that the CMB can be described as a single-source plane. The theoretical CMB lensing potential power spectrum $C^{\phi\phi}_\ell$. can be expressed as \cite{Lewis:2006fu,Hanson:2009kr}
\be
\label{eq:CL_phiphi}
C_\ell^{\phi\phi} = 16 \pi \int \frac{dk}{k}
P_{\R}(k) \left[ \int_0^{\chi_*} d\chi T^\lin_{\phi}(k; \eta_0 - \chi) j_\ell(k\chi)\left( \frac{\chi_* - \chi}{\chi_* \chi} \right) \right]^2 \, .
\ee
In this expression, $\chi$ denotes the comoving distance, while $\eta$ denotes the conformal time and $j_\ell$ is the spherical Bessel function. The line-of-sight integral $\int_0^{\chi_*}d\chi$ runs from the observer ($\chi=0$) to the CMB last scattering surface ($\chi=\chi_*)$; $P_\R$ and $T^\lin_{\phi}$ denote the power spectrum of primordial curvature perturbations $\R$ and the transfer function in linear theory, respectively. That is,
\be
\label{eq:lin_transfer}
\phi(k,\eta)=T^\lin_{\phi}(k,\eta)\R(k).
\ee
In this work, we scale the linear transfer function by the same growth function in Eqs.(3)-(4) of the Letter as
\be
\label{eq:lin_transfer_gamma}
T^\lin_{\phi}(\gamma,k,a)\to T^\lin_{\phi}(k,a)\,\frac{D(\gamma,a)}{D(\gamma=0.55,a)}.
\ee
Nonlinear evolutions of $\phi$, hence enhancement of $C_\ell^{\phi\phi}$, can be approximated by further scaling $T^\lin_{\phi}$ according to
\be
\label{eq:nonlin_transfer}
T^\nonlin_{\phi}(\gamma,k,a)=T^\lin_{\phi}(\gamma,k,a)\left[\frac{P^\nonlin(\gamma,k,a)}{P^\lin(\gamma,k,a)}\right]^{1/2}.
\ee
This nonlinear approximation is assumed by the fiducial \code{CAMB} code \cite{Lewis:camb,Howlett:camb} and PL18 analyses \cite{Planck2018:cosmology,Planck2018:lensing} which we follow here.
Note that the linear and nonlinear matter power spectra are given by Eq.~(4) of the Letter and \refeq{nonlin_pk} of the Supplementary Material, respectively.

In \reffig{lensing_phiphi_Dell_freegamma_vs_fixedgamma_PL18_temp-pol_bestfit}, we show the theoretical $C_\ell^{\phi\phi}$ predicted for different cosmologies: one with $\gamma=0.55$ and one with free $\gamma$ that best fit the PL18 temp.+pol. data. We additionally plot their observational counterpart: the PL18 lensing reconstruction band power $C_L^{\phi\phi}$.
The model with higher $\gamma$ enhances $C_\ell^{\phi\phi}$ in the multipole range of $L\sim10-10^3$, providing a better fit to $C_L^{\phi\phi}$ below $L\simeq100$.
This agrees with Fig.~2 of the Letter, as low-redshift gravitational potentials mostly contribute towards lensing of CMB on large scales, hence the more prominent difference between the two predictions at low $L$. We note that, in this particular case, even though the primary CMB spectra effectively fixes the amplitude $A_s$ of the primordial power spectrum $P_\R(k)$, the CMB lensing effect---which determines the smoothing of the CMB spectra---also probes $P(k,z)$ at low redshift, and for PL18 TTTEEE spectra, prefers \emph{higher} $P(k,z)$ amplitude or effectively, higher $D(\gamma,z)$ and $\gamma>0.55$.
Further, the ratio between two lensing spectra is compatible with the best-fit $A_{\mathrm{lens}}$ in the flat model with $\gamma=0.55$ investigated in the main text. Both are also consistent with $A_{\mathrm{lens}}=1.18\pm0.065$ inferred from PL18 temp.+pol. in \cite{Planck2018:cosmology}.

On the other hand, as also noted in \cite{Planck2018:cosmology}, the PL18 reconstructed $C_L^{\phi\phi}$ prefers spectra that are slightly tilted towards less power at high $L$. This explains the positive $\Delta\chi^2$ for the ``lensing reconstruction'' data and likelihood in \reftab{individual_delta_chisquares}. In other words, PL18 CMB lensing reconstruction by itself does not favor high $\gamma$. It is the smoothing effect of PL18 CMB temperature and polarization power spectra that prefers $\gamma>0.55$.

\begin{figure}[t]
   \centering
 \includegraphics[width=\linewidth]{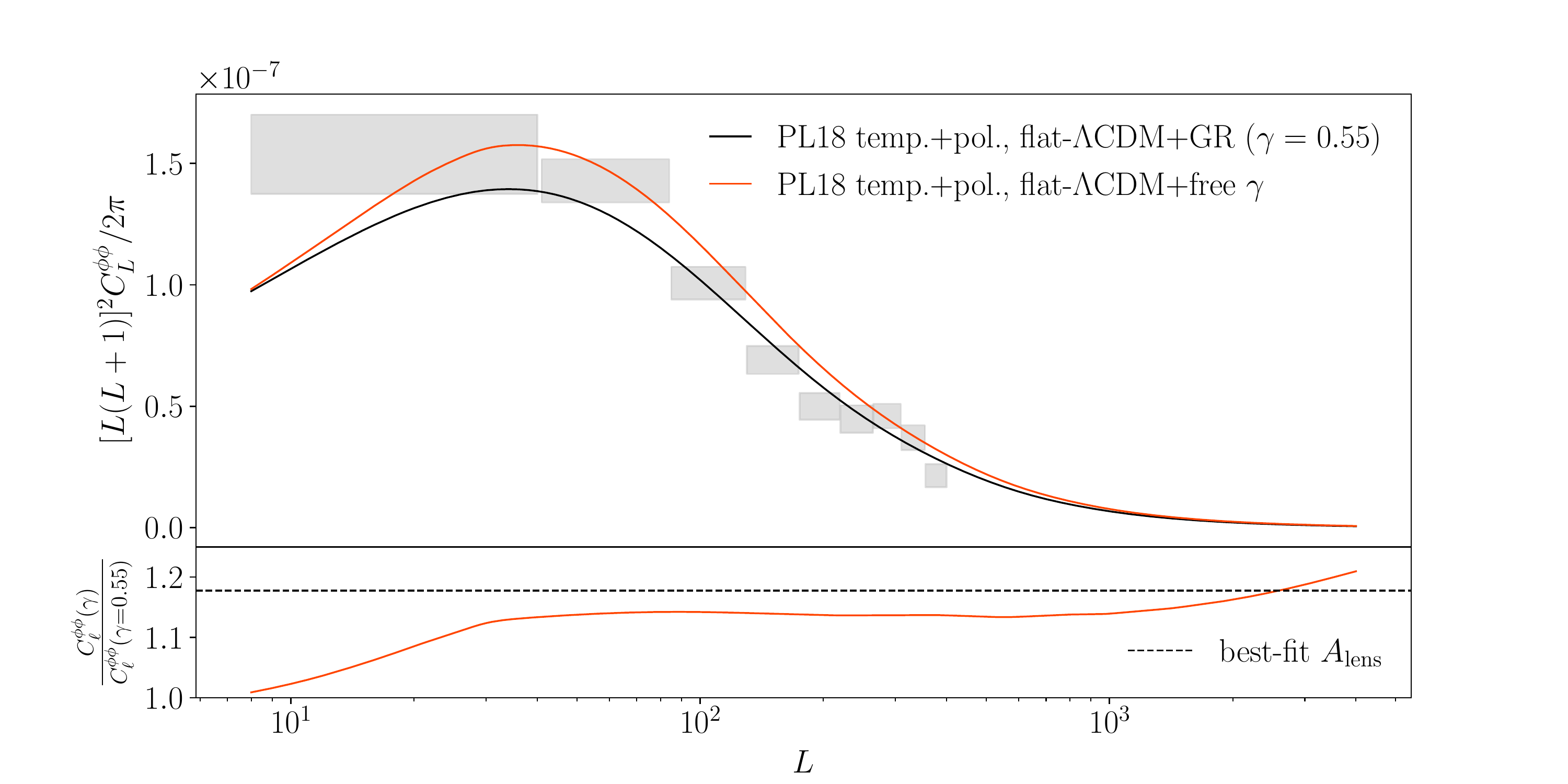}
   \caption{Top panel: The CMB lensing power spectra $C^{\phi\phi}_L$ in PL18 temp.+pol. best-fit models with $\gamma=0.55$ (black) and free $\gamma$ (orange). The grey boxes indicate the Planck 2018 estimates of the CMB lensing bandpowers and their uncertainties. Bottom panel: The ratio between the two predicted lensing power spectra. Horizontal line indicates the best-fit $A_{\mathrm{lens}}$ for the same data, at $\gamma=0.55$.}
   \label{fig:lensing_phiphi_Dell_freegamma_vs_fixedgamma_PL18_temp-pol_bestfit}
 \end{figure}

\section{\texorpdfstring{IV.\hspace{0.05in}Consistency and Robustness of $\gamma$ constraints}{Consistency and Robustness of gamma constraints}}\label{subsec:robust-fsigma8}

\begin{figure}[htp]
\centering
\includegraphics[width=.33\textwidth]{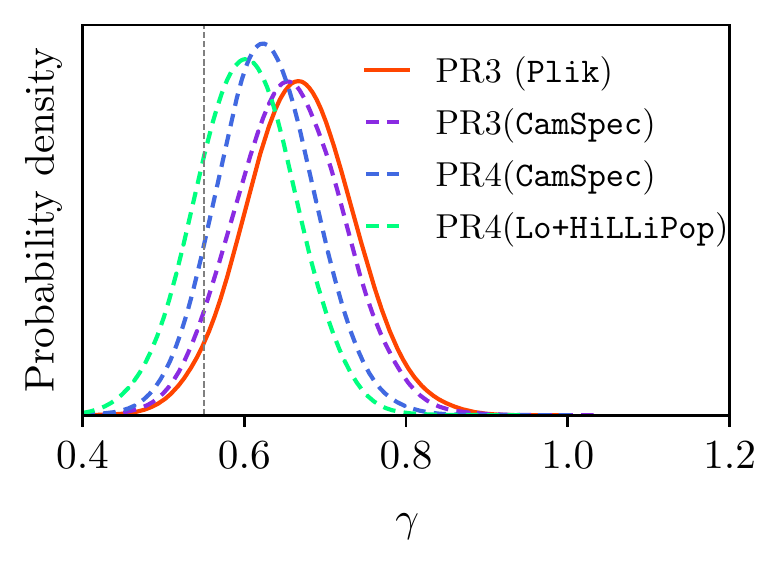}\hfill
\includegraphics[width=.33\textwidth]{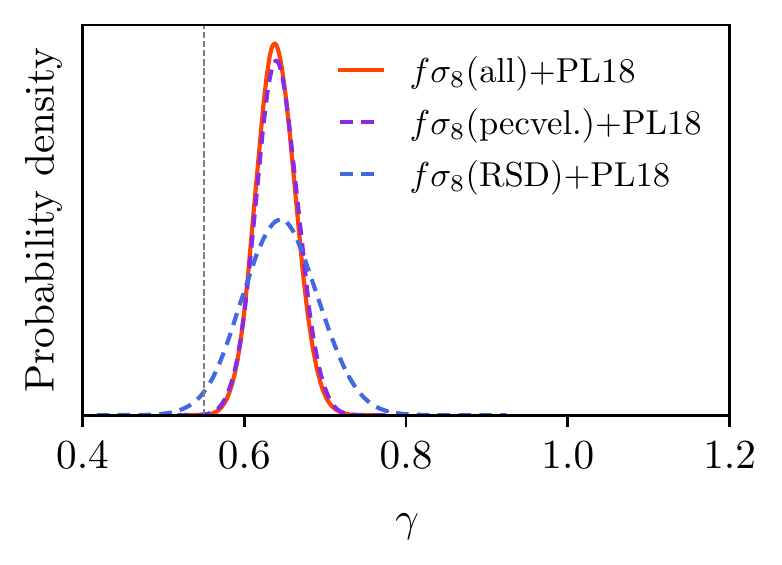}\hfill
\includegraphics[width=.33\textwidth]{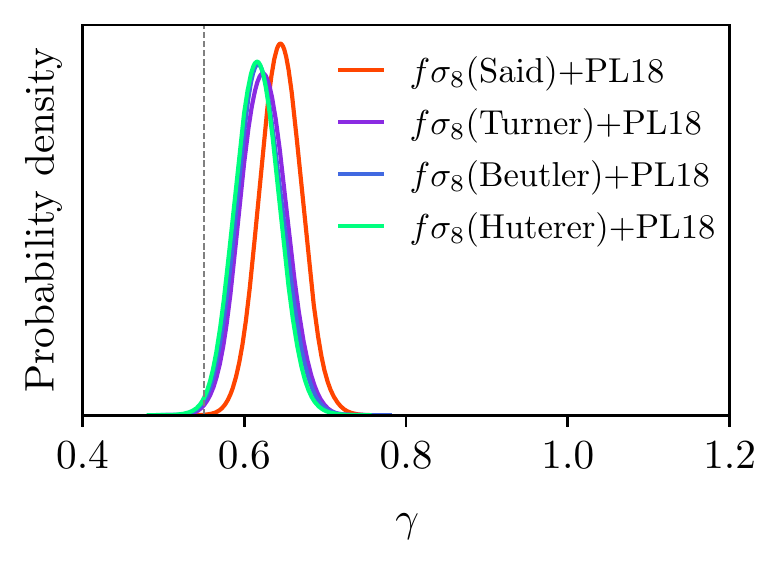}
\caption{Marginalized constraints on $\gamma$ obtained from different data combinations. Left panel: $\gamma$ constraints from different Planck data and likelihood releases. Middle panel: $\gamma$ constraints from PL18 (baseline) and $f\sigma_8$ measurements using peculiar velocity, or RSD, or both (baseline). Right panel: $\gamma$ constraints from PL18 and $f\sigma_8$ measurements, where the latter include only measurements that are \emph{uncorrelated}; labels indicate which one of the potentially correlated data points from \cite{Beutler:2012px,Huterer:2016,Said:2020,Turner:2023} is being used (see text).}
\label{fig:robust_gamma}
\end{figure}

\textbf{Planck likelihoods and data releases.}
Our CMB data sets and likelihoods, PL18, are taken from the latest official Planck cosmological parameter analysis (Planck PR3) \cite{Planck2018:cosmology}. Since that result, there have been new releases of a) Planck maps---specifically the Planck PR4 (\code{NPIPE}) maps where high- and low-frequency maps were jointly processed using the same pipeline \cite{Planck:2020olo}---and b) low+high-$\ell$ polarization-based likelihoods \cite{Tristram:2020wbi} or the updated \code{CamSpec} high-$\ell$ likelihood \cite{Efstathiou:2019mdh,Rosenberg:2022sdy}, both based on the PR4 \code{NPIPE} maps. In addition, there is a new CMB lensing reconstruction from these new maps \cite{Carron:2022eyg}. Notably, there have been significant shifts in $\Omega_K$ and $A_{\mathrm{lens}}$ reported by the new analyses.
To assert the robustness of our $\gamma$ constraint with respect to Planck data, in the left panel of \reffig{robust_gamma} we compare $\gamma$ constraints from each of these new sets of Planck maps and likelihoods. Here, ``PR3 (\code{Plik})'' represents the constraint from PR3 maps and likelihoods---as used in our analysis; ``PR3(\code{CamSpec})'' and ``PR4(\code{CamSpec})'' represent the constraints obtained using the high-$\ell$ \code{CamSpec} likelihood from PR3 maps \cite{Efstathiou:2019mdh} and PR4 \code{NPIPE} maps \cite{Rosenberg:2022sdy}, respectively; ``PR4 (\code{Lo+HiLLiPop})'' represents the constraint from the PR4 maps and \code{LoLLiPop}+\code{HiLLiPop} likelihoods for low+high-$\ell$ EEEBBB+TTTEEE spectra \cite{Planck:2020olo,Tristram:2020wbi}. We find that all variations of Planck analyses are consistent in their constraints of $\gamma$. For lensing reconstruction, ``PR3'' labels imply the PR3 lensing reconstruction while ``PR4'' labels indicate the new PR4 \code{NPIPE} lensing reconstruction \cite{Carron:2022eyg}. Switching entirely from PR3 to PR4 data and likelihoods result in a slight decrease in the significance for $\gamma>0.55$, going from 1.9-2$\sigma$ with PR3 (\code{CamSpec}-\code{Plik}) to 1.2-1.6$\sigma$ with PR4 (\code{Lo+HiLLiPop}-\code{CamSpec}). We note that the most significant shift occurs with ``PR4 (\code{Lo+HiLLiPop})'' which involves low-$\ell$ spectra from \code{NPIPE} polarization maps. These are known to require careful calibration of their low-$\ell$ transfer function, due to new map calibration scheme \cite{Planck:2020olo}; moreover, even their high-$\ell$ EE spectra might be prone to systematics as discussed in \cite{Rosenberg:2022sdy}.

\medskip
\textbf{RSD and peculiar-velocity measurements of $f\sigma_8$}.
The $f\sigma_8$ data we use are extracted from both RSD and peculiar-velocity measurements, which are typically affected by different systematics. Before combining those data points, we therefore examine the consistency between their preferences for $\gamma$. In the center panel of \reffig{robust_gamma}, we show $\gamma$ constraints from $f\sigma_8$ data measured by either RSD or peculiar velocities, in both cases combined with Planck CMB data. We thereby verify that the constraints from $f\sigma_8$(pecvel.)+PL18 (dashed violet) and $f\sigma_8$(RSD)+PL18 (dashed blue) are mutually consistent: both prefer $\gamma>0.55$. We further notice that peculiar-velocity measurements of $f\sigma_8$ significantly contribute towards the baseline constraint $f\sigma_8$(all)+PL18 (solid orange), more so than the RSD counterparts.

\section{\texorpdfstring{V.\hspace{0.05in}Unaccounted correlation between $f\sigma8$ measurements}{Unaccounted correlation between fsigma8 measurements}}\label{subsec:corrrelated-fsigma8}
Some of our $f\sigma_8$ data points in Fig.~2 of the Letter were measured in the same or overlapping survey(s), though using different (sub-)samples or methods. These include the peculiar-velocity or RSD measurements in \cite{Beutler:2012px,Huterer:2016,Said:2020,Turner:2023}.
Could our $\gamma$ constraints significantly underestimate the uncertainties by ignoring the possible correlation between those measurements? Explicitly accounting for the associated non-diagonal terms in the covariance for $f\sigma_8$ data is highly nontrivial. To make progress and address the question, we repeat our analysis by keeping only \emph{one} of the measurements in \cite{Beutler:2012px,Huterer:2016,Said:2020,Turner:2023} at a time, dropping the other three data points from the baseline $f\sigma_8$ data in Fig.~2 of the Letter.
We show the results in the right panel of \reffig{robust_gamma}, where the legend indicates which of the $f\sigma_8$ measurements in \cite{Beutler:2012px,Huterer:2016,Said:2020,Turner:2023} we use. All constraints agree with our fiducial constraint, and show that we do not significantly overestimate the significance of $\gamma>0.55$. 

\end{document}